# The Capacity of the Semi-Deterministic Cognitive Interference Channel and its Application to Constant Gap Results for the Gaussian Channel

Stefano Rini, Daniela Tuninetti, and Natasha Devroye
Department of Electrical and Computer Engineering
University of Illinois at Chicago
Email: {srini2, danielat, devroye}@uic.edu

*Abstract*—The cognitive interference channel (C-IFC) consists of a classical two-user interference channel in which the message of one user (the "primary" user) is non-causally available at the transmitter of the other user (the "cognitive" user). We obtain the capacity of the semi-deterministic C-IFC: a discrete memoryless C-IFC in which the cognitive receiver output is a noise-less deterministic function of the channel inputs. We then use the insights obtained from the capacity-achieving scheme for the semi-deterministic model to derive new, unified and tighter constant gap results for the complex-valued Gaussian C-IFC. We prove: (1) a constant *additive gap* (difference between inner and outer bounds) of half a bit/sec/Hz per real dimension, of relevance at high SNRs, and (b) a constant *multiplicative gap* (ratio between outer and inner bounds) of a factor two, of relevance at low SNRs.

## I. INTRODUCTION

The Cognitive InterFerence Channel (C-IFC) consists of a classical two-user interference channel with two independent messages, in which the messages of the "primary" user is non-causally provided to the transmitter of the other "secondary" or "cognitive" user. This channel models the ability of intelligent and adaptive devices to listen to the wireless environment and obtain information about the other nodes' activity, which may be utilized in a selfish or altruistic manner in the transmission of their own message. The C-IFC may be seen as a limiting case of the cooperative communications paradigm where cooperation between transmitters is modeled as *asymmetric* and *non-causal* – an idealization of the more realistic causal cooperation.

The capacity of the C-IFC is known only for some classes of Discrete Memoryless (DM) and Additive White Gaussian Noise (AWGN) channels. In particular, capacity is known for the linear high-SNR deterministic C-IFC [1], a deterministic channel that models the AWGN C-IFC in the high-SNR regime, and for certain AWGN C-IFCs [2]–[7]. In this paper we derive the capacity of the general semi-deterministic C-IFC: a discrete memoryless C-IFC in which the cognitive output is a deterministic function of the channel inputs, while the primary output remains fully general (probabilistic). This

The work of S. Rini and D. Tuninetti was partially funded by NSF under award 0643954.

class of channels contains the linear high-SNR deterministic channel [1] as special case. We next use the intuition provided by the capacity achieving scheme of the semi-deterministic C-IFC to obtain an achievable rate region for the AWGN C-IFC which lies to within half a bit/sec/Hz per real dimension of the outer bound in [7, Th.4] – improving both on the previous available gap result of about two bits/sec/Hz per real dimension of [8], as well as achieving it with a single scheme. We furthermore obtain a multiplicative gap of a factor two for the AWGN C-IFC, of importance in approximately characterizing the channel capacity at low-SNR regime. This multiplicative gap result improves on the result in [6] as it holds for a general AWGN C-IFC.

### A. Past Work

The C-IFC was first introduced in [9] and has since been explored by numerous groups. Due to space constraints we outline only a few key results; more extended treatments and comparisons of achievable rate regions and outer bounds may be found in [5]. For the general DM C-IFC, capacity was known in the "very weak interference" regime [3], and in the "very strong interference" regime [4]; recently these two results have been unified and extended in [5] where capacity was shown for the "better cognitive decoding" regime. For the AWGN C-IFC, capacity is known in the "weak interference" regime [2], [3], in the "very strong interference" regime [4], and in the "primary decodes cognitive" regime [6].

Most of the past work focussed on DM and the AWGN C-IFCs. In this work we will first consider a class of semi-deterministic C-IFCs. Historically, deterministic channel models, where all channel outputs are a deterministic function of the inputs, have in some cases led to capacity results which are unknown for their random counterparts [10], [11]. Deterministic models have inspired the linear high-SNR deterministic approximation of AWGN channels in [12] by capturing the dynamic range of the desired and interfering signals while neglecting the additive noise. Determining the capacity of the (often significantly easier) deterministic approximation provides important insights on how to derive tighter inner and outer bounds for the original AWGN channel. Successful examples of this approach include the AWGN interference channel (with [13] and without [14] feedback), the relay channel [12], and the C-IFC [1]. In particular, in [15] we obtained the capacity of the linear high-SNR deterministic C-IFC and we used it in [8] to prove a constant additive gap capacity result of 1.87 bits/sec/Hz per real dimension for the general AWGN C-IFC.

In [16] a class of semi-deterministic DM C-IFCs is introduced, in which the cognitive output is a deterministic function of the channel inputs, and capacity is derived under the assumptions that the deterministic function satisfies an invertibility condition and a mutual information constraint equivalent to the "better cognitive decoding" regime [5].

*B. Main Contributions*

In this work we extend previous results on the semi-deterministic DM and AWGN C-IFC in a number of ways:
1) We obtain the capacity region of the semi-deterministic DM C-IFC without any further constraints, thus extending the result of [16].
2) We use the insight obtained from the capacity-achieving scheme for the semi-deterministic DM C-IFC to devise an achievable scheme for the AWGN C-IFC. We show that this scheme achieves a constant additive gap of a half bit/sec/Hz per real dimension. This extends our previous result in [8] in two ways: a) we reduce the gap from 1.87 bits to half a bit per real-dimension; and b) the proof uses a single achievable region for all possible channel parameters, rather than different schemes for different parameter regimes.
3) We show a multiplicative gap of a factor two for the general AWGN C-IFC, improving and extending our multiplicative gap result of [6] which holds only for a subset of the strong interference regime.

*C. Paper Organization*

In Section II we introduce the channel model, in Section III we derive the capacity for a class of semi-deterministic DM C-IFCs, and in Section IV we derive additive and multiplicative gaps for the AWGN C-IFC. Section V concludes the paper.

## II. CHANNEL MODEL

*A. General memoryless cognitive interference channel*

A two-user C-IFC [9] is a multi-terminal network with two input alphabets $\mathcal{X}_1$ and $\mathcal{X}_2$, two output alphabets $\mathcal{Y}_1$ and $\mathcal{Y}_2$, and a channel transition probability $P_{Y_1Y_2|X_1X_2}(y_1,y_2|x_1,x_2) : \mathcal{Y}_1 \times \mathcal{Y}_2 \to [0,1]$ for all $(x_1,x_2) \in \mathcal{X}_1 \times \mathcal{X}_2$. Each transmitter $i$, $i \in \{1,2\}$, wishes to communicate a message $W_i$, uniformly distributed on $[1:2^{NR_i}]$, to receiver $i$ in $N$ channel uses at rate $R_i$ bis per channel use. The two messages are independent. Transmitter 1, in addition to its own message $W_1$, also knows the message $W_2$ of transmitter 2. Transmitter/receiver 1 is referred to as the *cognitive* pair and transmitter/receiver 2 as the *primary* pair. A graphical representation of a C-IFC is found in Fig. 1. Achievable rates and capacity region are defined in the usual way [17].

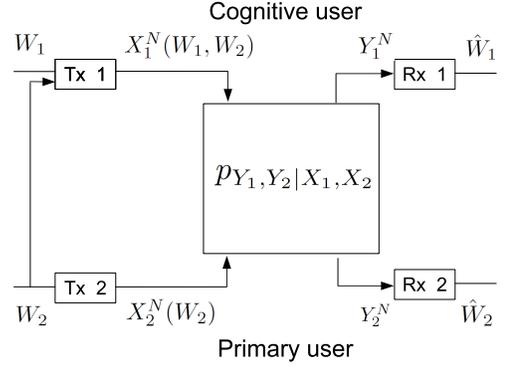

Fig. 1. The C-IFC model.

The C-IFC is an idealized model for the *unilateral source cooperation* of transmitter 1 with transmitter 2. The receivers however do not cooperate. This implies that the capacity region of the C-IFC, similarly to the broadcast channel (BC) [18], only depends on the output conditional marginals $P_{Y_1|X_1X_2}$ and $P_{Y_2|X_1X_2}$, and not on the output joint marginal $P_{Y_1Y_2|X_1X_2}$. This important observation is useful in deriving sum-rate outer bounds.

*B. Semi-deterministic C-IFC*

We start by considering a class of semi-deterministic C-IFCs for which the received signal at the cognitive receiver (receiver 1) is a deterministic function of the channel inputs,

$$Y_1 = f_1(X_1, X_2), \qquad (1)$$

for some function $f_1 : \mathcal{X}_1 \times \mathcal{X}_2 \to \mathcal{Y}_1$.

*C. AWGN C-IFC*

An AWGN C-IFC (G-C-IFC) in *standard form* has outputs

$$Y_1 = X_1 + aX_2 + Z_1,$$
$$Y_2 = |b|X_1 + X_2 + Z_2,$$

where the channel gains $a$ and $b$ are complex-valued, constant and known to all terminals, the channel inputs are subject to the power constraint

$$\mathrm{E}[|X_i|^2] \leq P_i, \qquad P_i \in \mathbb{R}^+, \qquad i \in \{1,2\},$$

and the channel noise $Z_i \sim \mathcal{N}_{\mathbb{C}}(0,1)$, $i \in \{1,2\}$. Since the capacity only depends on the output conditional marginals [5], the correlation coefficient of $Z_1$ and $Z_2$ is irrelevant.

## III. CAPACITY FOR THE SEMI-DETERMINISTIC C-IFC

We now obtain the capacity of the semi-deterministic C-IFC in (1). This class of channels was first considered in [16] where the capacity region was derived under the additional assumptions: (a) $I(Y_1;X_2) \geq I(Y_2;X_2)$ for all possible input distributions, and (b) $f_1$ in (1) is invertible. Here we extend the result of [16] *without requiring conditions (a) and (b)*.

**Theorem 1.** *The capacity region of the semi-deterministic DM C-IFC in (1) consists of all $(R_1, R_2) \in \mathbb{R}_+^2$ such that*

$$R_1 \leq H(Y_1|X_2), \tag{2a}$$
$$R_2 \leq I(Y_2; U, X_2), \tag{2b}$$
$$R_1 + R_2 \leq I(Y_2; U, X_2) + H(Y_1|U, X_2), \tag{2c}$$

*taken over the union of all distributions $P_{UX_1X_2}$.*

*Proof:*
*Converse:* In [3, Th.3.2] it is shown that the capacity region of a DM C-IFC is contained in:

$$R_1 \leq I(Y_1; X_1|X_2), \tag{3a}$$
$$R_2 \leq I(Y_2; U, X_2), \tag{3b}$$
$$R_1 + R_2 \leq I(Y_2; U, X_2) + I(Y_1; X_1|X_2, U), \tag{3c}$$

for some $P_{UX_1X_2}$. The converse follows by evaluation the constraints in (3a) and in (3c) under the deterministic assumption in (1) i.e. $H(Y_1|X_1, X_2) = 0$.

*Achievability:* The largest known achievable region of [**?**, Th.5.1]RTDjournal1
computed for $U_{2c} = U_{1c} = \emptyset$, $U_1 = U_{1pb}$ and $U_2 = U_{2pb}$, and $R_{2c} = R_{1c} = R_{1pb} = 0$ becomes:

$$R_1' \geq I(U_1; X_2), \tag{4a}$$
$$R_1' + R_2' \geq I(U_1; U_2, X_2), \tag{4b}$$
$$R_2 + R_2' \leq I(Y_2; U_2, X_2), \tag{4c}$$
$$R_1 + R_1' \leq I(Y_1; U_1), \tag{4d}$$

for any $P_{U_1U_2X_1X_2}$. After Fourier-Motzkin elimination, the region in (4) may be rewritten as

$$R_1 \leq I(Y_1; U_1) - I(U_1; X_2), \tag{5a}$$
$$R_2 \leq I(Y_2; U_2, X_2)$$
$$= I(Y_2; X_1, X_2) - I(Y_2; X_1|U_2, X_2), \tag{5b}$$
$$R_1 + R_2 \leq I(Y_2; U_2, X_2) + I(Y_1; U_1) - I(U_1; U_2, X_2)$$
$$= (5a) + (5b) - I(U_1; U_2|X_2). \tag{5c}$$

Finally, by choosing $U_1 = Y_1$ (possible because $Y_1$ is a deterministic function of the inputs and both inputs are known at transmitter 1) and $U_2 = U$, the achievable region in (5) reduces to the outer bound. ∎

**Theorem 2.** *If both outputs are deterministic functions of the channel inputs, that is*

$$Y_1 = f_1(X_1, X_2), \ Y_2 = f_2(X_1, X_2), \tag{6}$$

*then the capacity region is given by*

$$R_1 \leq H(Y_1|X_2), \tag{7a}$$
$$R_2 \leq H(Y_2), \tag{7b}$$
$$R_1 + R_2 \leq H(Y_2) + H(Y_1|Y_2, X_2), \tag{7c}$$

*taken over the union of all distributions $p_{X_1,X_2}$.*

*Proof:* A deterministic channel belongs to the class of semi-deterministic channels whose capacity is given by Th.1. The achievability of (7) thus follows immediately by choosing $U = Y_2$ in the capacity region (2). Note that it is possible to set $U = Y_2$ because the codebook $U$ is generated at the cognitive transmitter that knows both inputs (and thus knows $Y_2$ because $Y_2$ is a deterministic function of the inputs by assumption.)

The choice $U = Y_2$ also maximizes the $R_2$-bound in (2b) since

$$I(Y_2; U, X_2) \leq I(Y_2; U, X_1, X_2) = H(Y_2),$$

where the last equality follows from assumption (6); however, it is not evident a priori that it also maximizes the sum-rate in (2c). To show that the sum-rate is indeed bounded by (7c), we use the following sum-rate outer bound which can be developed following steps similar to those in [19]:

$$R_1 + R_2 \leq I(X_1, X_2; Y_2) + I(Y_1; X_1|Y_2', X_2),$$

where $Y_2'$ has the same conditional marginal distribution as $Y_2$. Since we are dealing with deterministic channels, we can only choose $Y_2' = Y_2$, from which the claim follows. ∎

**Remark 1.** *The achievability part of Th.2 shows that for the deterministic C-IFC the choice $U_1 = Y_1$ and $U_2 = Y_2$ in (5) is optimal. We will use this observation in the next section.*

IV. CAPACITY RESULTS FOR THE AWGN C-IFC

We next determine constant bounds for the capacity of the AWGN C-IFC. Specifically, we bound the "distance" between the inner bound in (5) and the known outer bound for the AWGN C-IFC in [4]. We use two "distance" measures: the additive gap and the multiplicative gap. An additive gap is an upper bound on the difference between between outer and inner bounds which is valid for *all* channel parameters. Similarly, a multiplicative gap is an upper bound on the ratio between outer and inner bounds which is valid for *all* channel parameters.

*A. Additive Gap*

**Theorem 3.** *For a AWGN C-IFC, the capacity is known to within half a bit/sec/Hz per real dimension (or one bit/sec/Hz for the complex-valued channel).*

*Proof:* The capacity for weak interference ($|b| \leq 1$) was determined in [3], so we only need to concentrate on the strong interference regime ($|b| > 1$).

*Outer bound:* In [7, Th. 4] it was shown that the capacity region of a AWGN C-IFC with strong interference lies in

$$R_1 \leq \log(1 + \alpha P_1), \tag{8a}$$
$$R_1 + R_2 \leq \log\left(1 + |b|^2 P_1 + P_2 + 2\sqrt{\bar{\alpha}|b|^2 P_1 P_2}\right), \tag{8b}$$

taken over the union of all $\alpha \in [0, 1]$, with $\bar{\alpha} = 1 - \alpha$.

*Achievability to within 0.5 bit/sec/Hz per real dimension or 1 bit/sec/Hz per complex dimension:* A constant additive gap is shown using the region in (5) with the following choice of

random variables:

$$X_{1pb} \sim \mathcal{N}_\mathbb{C}(0, \alpha P_1) \tag{9a}$$
$$X_2 \sim \mathcal{N}_\mathbb{C}(0, P_2), \text{ independent of } X_{1pb}, \tag{9b}$$
$$X_1 = X_{1pb} + \sqrt{\frac{\bar{\alpha} P_1}{P_2}} X_2 \tag{9c}$$
$$U_1 = X_1 + aX_2 + Z_{1pb} \tag{9d}$$
$$U_2 = |b|X_1 + X_2 + Z_{2pb}, \tag{9e}$$

where

$$\begin{bmatrix} Z_{1pb} \\ Z_{2pb} \end{bmatrix} \sim \mathcal{N}_\mathbb{C} \left( 0, \begin{bmatrix} \sigma_{1pb}^2 & \rho_{pb} \sqrt{\sigma_{1pb}^2 \sigma_{2pb}^2} \\ \rho_{pb}^* \sqrt{\sigma_{1pb}^2 \sigma_{2pb}^2} & \sigma_{2pb}^2 \end{bmatrix} \right),$$

for $|\rho_{pb}| \leq 1$. The assignment proposed in (9) is inspired by the capacity achieving scheme for deterministic channels in Th. 2, where we showed that $U_c = Y_c$, $c \in \{1,2\}$, is optimal. In a noisy channel, it is not possible to choose $U_c = Y_c$; we mimic this by setting $U_c \sim Y_c$, $c \in \{1,2\}$.

With the assignment (9), the following region is achievable:

$$R_1 \leq \log(\sigma_{1pb}^2 + \alpha P_1)$$
$$- \log\left(\sigma_{1pb}^2 + \frac{\text{Var}[X_1 + aX_2]}{1 + \text{Var}[X_1 + aX_2]}\right), \tag{10a}$$
$$R_2 \leq \log(1 + \text{Var}[|b|X_1 + X_2])$$
$$- \log\left(1 + \frac{\sigma_{2pb}^2 |b|^2 \alpha P_1}{\sigma_{2pb}^2 + |b|^2 \alpha P_1}\right), \tag{10b}$$
$$R_1 + R_2 \leq (10a) + (10b)$$
$$+ \log\left(1 - \frac{\left|[|b|P_1 \alpha - \sqrt{\sigma_{1pb}^2 \sigma_{2pb}^2}]^+\right|^2}{(|b|^2 P_1 \alpha + \sigma_{2pb}^2)(P_1 \alpha + \sigma_{1pb}^2)}\right), \tag{10c}$$

where $[x]^+ = \max\{0, x\}$ and

$$\text{Var}[X_1 + aX_2] = P_1 + |a|^2 P_2 + 2\text{Re}\{a\}\sqrt{\bar{\alpha} P_1 P_2},$$
$$\text{Var}[|b|X_1 + X_2] = |b|^2 P_1 + P_2 + 2\sqrt{\bar{\alpha}|b|^2 P_1 P_2}.$$

The RHS of (10c) follows by choosing

$$\rho_{pb} = \arg\min I(U_1; U_2 | X_2)$$
$$= \arg\min |\mathbb{E}[U_1 U_2^* | X_2]|^2$$
$$= \arg\min \left||b|P_1 \alpha + \rho_{pb}\sqrt{\sigma_{1pb}^2 \sigma_{2pb}^2}\right|^2$$
$$= -\min\left\{1, \frac{|b|P_1 \alpha}{\sqrt{\sigma_{1pb}^2 \sigma_{2pb}^2}}\right\}.$$

With $\sigma_{2pb}^2 = 0$ and $\sigma_{1pb}^2 = 1$ in (10) we have

$$R_1 \leq \log(1 + \alpha P_1) - \text{GAP}(\alpha), \tag{11a}$$
$$R_1 + R_2 \leq \log(1 + \text{Var}[|b|X_1 + X_2]) - \text{GAP}(\alpha), \tag{11b}$$

with $\text{GAP}(\alpha)$ bounded as

$$\text{GAP}(\alpha) = \log\left(1 + \frac{\text{Var}[X_1 + aX_2]}{1 + \text{Var}[X_1 + aX_2]}\right) \leq \log(2) = 1,$$

as claimed. Notice that with $\sigma_{2pb}^2 = 0$, the $R_2$-bound in (10b) is equivalent to the sum-rate outer bound in (8b) and it is thus redundant. ∎

**Remark 2.** *The additive gap result in Th. 3 improves on our previous result in [8] in a number of ways: (1) it uses a single scheme to achieve a constant gap in the whole parameter space (while in [8] we used two different schemes for different parameter regimes), (2) the present result is valid for complex-valued channels (while in [8] we only considered real-valued channels), and (3) the gap is reduced from 1.87 bits per real-dimension in [8] to half a bit per real-dimension (since one complex-dimension is equivalent to two real-dimensions).*

Notice that $\sigma_{1pb}$ and $\sigma_{2pb}$ could be optimized so as to get the smallest possible gap for all choice of parameters, rather than setting them to $\sigma_{1pb} = 1$, $\sigma_{2pb} = 0$.

*B. Multiplicative Gap*

**Theorem 4.** *For a AWGN C-IFC, the capacity is known to within a factor two.*

*Proof:* The capacity for weak interference ($|b| \leq 1$) was determined in [3], thus we only need to concentrate on the strong interference regime ($|b| > 1$).

*Outer bound:* We again use the outer bound in (8), which we rewrite as

$$R_2 \leq \log\left(1 + |b|^2 P_1 + P_2 + 2\sqrt{|b|^2(1 + P_1 - e^{R_1})P_2}\right) - R_1$$
$$\triangleq R_2^{(\text{outer bound})}(R_1), \tag{12}$$

for $R_1 \in [0, \log(1 + P_1)]$.

*Achievability to within a factor two:* Consider the following TDMA strategy. The rate-point

$$(R_1, R_2) = (\log(1 + P_1), 0),$$

is achievable by silencing the primary transmitter, while the rate-point

$$(R_1, R_2) = \left(0, \log(1 + (\sqrt{|b|^2 P_1} + \sqrt{P_2})^2)\right),$$

is achievable by having both transmitters beamform to the primary receiver. Hence, the following region is achievable by time sharing

$$R_2 \leq \left(1 - \frac{R_1}{\log(1 + P_1)}\right) \log(1 + (\sqrt{|b|^2 P_1} + \sqrt{P_2})^2)$$
$$\triangleq R_2^{(\text{tdma})}(R_1). \tag{13}$$

Finally, the multiplicative gap is given by the smallest $M \geq 1$ for which

$$MR_2^{(\text{tdma})}(R_1/M) \geq R_2^{(\text{outer bound})}(R_1), \tag{14}$$

that is

$$M\left(1 - \frac{R_1}{M\log(1 + P_1)}\right) \log(1 + (\sqrt{|b|^2 P_1} + \sqrt{P_2})^2) \geq$$
$$\log\left(1 + |b|^2 P_1 + P_2 + 2\sqrt{|b|^2(1 + P_1 - e^{R_1})P_2}\right) - R_1, \tag{15}$$

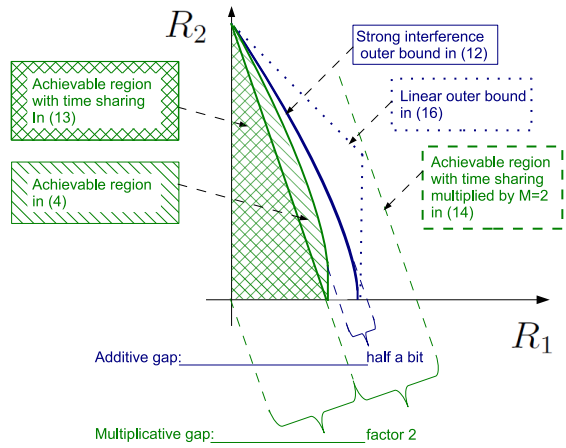

Fig. 2. A graphical representation of Th. 3 and Th. 4

for all $R_1 \in [0, \log(1+P_1)]$. By upper bounding the RHS of (15) so as to obtain an expression linear in $R_1$, we have

$$\left(1 - \frac{R_1}{M\log(1+P_1)}\right) M\log(1+(\sqrt{|b|^2 P_1}+\sqrt{P_2})^2)$$
$$- \log\left(1+(\sqrt{|b|^2 P_1}+\sqrt{P_2})^2\right) + R_1 \geq 0. \quad (16)$$

The LHS of (16) is a linear function of $R_1$ and thus has at most one zero. From this, it follows that the inequality in (16) is verified for every $R_1 \in [0, \log(1+P_1)]$ if it is verified in the boundary points of the interval. For $R_1 = 0$, the inequality is verified for $M \geq 1$ while for $R_1 = \log(1+P_1)$ it is verified if $M \geq 2$; thus the smallest $M$ for which (16) is verified is $M = 2$. ∎

A schematic plot of the proof of Th. 3 and Th. 4 is provided in Fig. 2. The green, hatched area represents the achievable region in (4) which lies to within half a bit/s/Hz from the outer bound of (8), illustrated by a solid blue line. The green, cross hatched area represents the achievable region with time sharing in (13) while the green dashed line the region in (13) multiplied by a factor two, which contains the linear outer bound in (16), illustrated by dotted blue line.

**Remark 3.** *The only previously available constant multiplicative gap result is that of [6] in which a gap of two is shown for a subset of the $|b| > 1$ regime. The result of Th. 4 extends the result of [6] so as to include the entire $|b| > 1$ regime.*

## V. CONCLUSIONS

We determined the capacity region of a class of discrete memoryless semi-deterministic cognitive interference channels where the output of the cognitive receiver is a deterministic function of the channel inputs. In this model, it is optimal for the cognitive encoder to perform joint binning as in Marton's achievability scheme for the broadcast channel. We then used the insights obtained form the capacity achieving scheme for the semi-deterministic channel to determine a novel approximate capacity results for the Gaussian channel: we determine the capacity of the Gaussian channel to within half a bit/sec/Hz and we complemented the additive gap (relevant at high SNR) with a multiplicative gap of a factor two of known outer bounds (relevant at low SNR).

Although the capacity region of the general C-IFC is still an open problem, ideas and techniques developed in this paper provide useful insights for the determination of novel inner and outer bounds; in particular, broadcast channel-like techniques appear, in several of the remaining open cases, to provide tighter bounds for the cognitive interference channel than similar interference channel-related techniques as showed by Th. 1, 2 and 3 in this paper.